\documentclass[letterpaper, 10 pt, conference]{ieeeconf}
\usepackage{enumerate}
\usepackage{courier}
\usepackage{hyperref}
\usepackage{graphicx,subfigure}
\usepackage{epstopdf}
\usepackage{tikz}
\usepackage{amsmath}
\usepackage{multirow}
\usepackage{amsfonts}
\usepackage{float}
\usepackage{amssymb}
\usepackage{graphicx}

\usepackage{flushend}
\usepackage{algorithm,float}
\usepackage{algpseudocode}
\usepackage{verbatim}
\newtheorem{theorem}{Theorem}
\newtheorem{lemma}{Lemma}

\newtheorem{remark}{Remark}
\newtheorem{example}{Example}

\makeatletter

\newcommand{\Rmnum}[1]{\expandafter\@slowromancap\romannumeral #1@}

\title{\LARGE \bf Emulation Learning for Neuromimetic Systems
}
\author{Zexin Sun \& John Baillieul}
\makeatother

\begin{document}
	\maketitle
	\thispagestyle{empty}
	\pagestyle{empty}

	\let\thefootnote\relax\footnotetext{\noindent\underbar{\hspace{0.8in}}\\
		Zexin Sun is with the Division of Systems Engineering at Boston University. John Baillieul is with the Departments of Mechanical Engineering, Electrical and Computer Engineering, and the Division of Systems Engineering at Boston University, Boston, MA 02215. The authors may be reached at {\tt \{zxsun, johnb\}@bu.edu}.\newline Support from various sources including the Office of Naval Research grant number N00014-19-1-2571 is gratefully acknowledged. }

\begin{abstract}
Building on our recent research on neural heuristic quantization systems, results on learning quantized motions and resilience to channel dropouts are reported. We propose a general emulation problem consistent with the neuromimetic paradigm. This optimal quantization problem can be solved by model predictive control (MPC), but because the optimization step involves integer programming, the approach suffers from combinatorial complexity when the number of input channels becomes large. Even if we collect data points to train a neural network simultaneously, collection of training data and the training itself are still time-consuming. Therefore, we propose a general Deep Q Network (DQN) algorithm that can not only learn the trajectory but also exhibit the advantages of resilience to channel dropout. Furthermore, to transfer the model to other emulation problems, a mapping-based transfer learning approach can be used directly on the current model to obtain the optimal direction for the new emulation problems.

\end{abstract}

	\section{Introduction}
	The work being reported continues our effort to develop the theoretical foundations of control system designs that exhibit key features of the neural mechanisms that govern movement and other behaviors in animals. One such feature is control modulation involving actions of very large numbers of simple inputs and outputs that are effective in influencing the system dynamics only in their aggregate operation. The goal of research on such control systems is to understand the engineering analogs of neuroplasticity, learning and relearning, memory and adaptation. Previously reported work has introduced what we call neuromimetic linear models. The focus has been on finite dimensional linear systems that are {\em overcomplete} in that they have many more input and output channels than the dimension of the state \cite{baillieul2014},\cite{baillieul2019}. Such systems have been shown to have reduced cost of operation in terms of standard $L_2$ metrics, reduced uncertainty in the face of input channel noise, and resilience with respect to drop-outs of input or output channels, \cite{baillieul2021neuromimetic}. With such qualities in mind, research is now aimed at overcomplete models with simple neuron-like discrete inputs taking values $\{-1,0,1\}$ and exploring how well these are able to emulate the behaviors of systems with standard continuous feedback designs. In \cite{Sun2022}, algorithms based on ideas from machine learning were used to solve a restricted emulation problem. The solutions were useful in comparing optimal and nearby suboptimal designs as well as in suggesting approaches to understanding the complexity of neuromimetic feedback designs.
	
	One of the approaches to feedback control using the quantized control inputs of \cite{Sun2022} involved reinforcement learning (RL)--specifically, a DQN-like algorithm \cite{DQN},\cite{Hessel2018}. RL can solve continuous decision-making tasks remarkably effectively. Agents can improve their performance by interacting with the environment through trial-and-error to minimize an emulation error metric \cite{Sutton2018}. However, when the state and action sets become large or possibly infinite, tabular solution methods must be abandoned in favor of approaches that lead only to approximate solutions. The approach here, following basic	ideas from \cite{DQN}, is to replace the Q-table of basic RL with a deep Q neural network (DQN). Revisiting \cite{Arulkumaran2017} in what follows, we examine the inherent complexity arising from the set of state-action pairs being infinite and the decision space being discrete. Even in this somewhat simple case, obtaining a sufficient sample of interactions as the emulation problem under study changes remains challenging since it is a time-sensitive problem, \cite{Zhu2020}. For this reason, we consider transfer learning (TL) \cite{Lazaric2012} as a way to utilize the experience from other emulation problems to accelerate the learning process for the new problem, \cite{Pan2009}.

	The rest of the paper is organized as follows. Section II describes the emulation problems, our system models, and the design goals of the learning algorithm. Section III introduces two approaches: an MPC-data based learning method and a generalized deep Q-network (DQN). Section IV presents the transfer learning algorithm to adapt the learned model a new emulation problem. Section V shows simulation results that illustrate the proposed approaches, and we make a summary in Section VI.
	
    \section{Problem Description}
\begin{figure}[h]
    	\begin{center}
    		\includegraphics[scale=0.5]{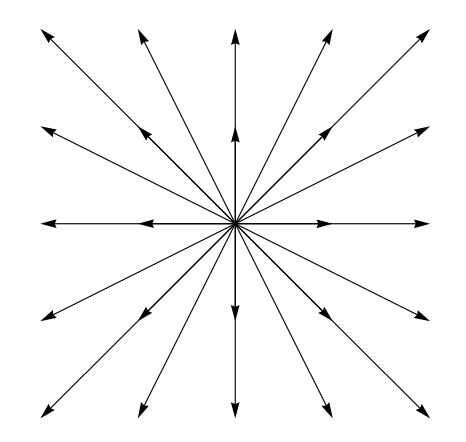}
    		\includegraphics[scale=0.6]{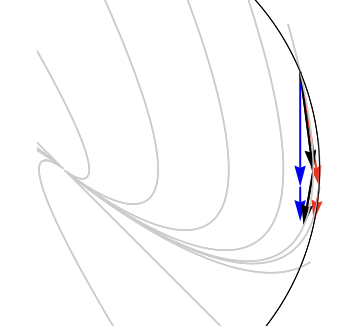}
    	\end{center}
    	\caption{The figure on the left is an example of a set of quantized directions (as considered in \cite{Sun2022} and related to the 81 activation patterns of Example 1), while the figure on the right shows choices of vector pairs (red vs blue) chosen for emulating a LTI vectorfield (gray curves and black tangents) when considering multiple steps.}
    	\label{fig:dir}
    \end{figure}

    	Consider the linear time-invariant (LTI) systems of the form 
    \begin{equation}
    	\begin{array}{l}
    		\dot x(t)=Ax(t) + Bu(t), \ \ \ x\in\mathbb{R}^n, \ \ u\in\mathbb{R}^m, \ {\rm and}\\[0.07in]
    		y(t)=Cx(t), \ \ \ \ \ \ \ \ \ \ \ \ \ \ y\in\mathbb{R}^q.\end{array}
    \end{equation} 
By applying the simple feedback control law $u=Kx$, where $K$ is a stabilizing gain chosen as in \cite{Sun2022}, we obtain the closed-loop LTI system
 \begin{equation}
    \dot x(t)=Hx(t),\ \ x(0)=x_0,\label{equation:LTI}
\end{equation} 
where $H=A+BK\in\mathbb{R}^{n\times n}$ is Hurwitz. Following \cite{baillieul2021neuromimetic}, \cite{Sun2022}, we consider the problem of emulating (\ref{equation:LTI}) by means of a discrete-time system with quantized inputs
    \begin{equation}
    	x_{qs}(k+1)=e^{Ah}x_{qs}(k)+\int_{0}^{h}e^{A(h-s)}Bu(k)ds,
    	\label{equation:qs}
    \end{equation}
    where $x_{qs}\in\mathbb{R}^n$ is the system state, $u\in\mathbb{U}=\{-1,0,1\}^m$ is the set of possible quantized inputs. Following \cite{Sun2022}, we refer to these inputs $u(k)$ as activation patterns. $h$ is the time step and $m\gg n$ denotes the large number of input channels. Here we provide an example.
    \begin{example}
    	Suppose the $n=2,m=4$ and that \[A=\begin{bmatrix}
    		0 & 0\\
    		0&0
    	\end{bmatrix}, B=\begin{bmatrix}
    	1 & 0&-1 & 0\\
    	0&1&0&1
    \end{bmatrix}.\] Then the update law (\ref{equation:qs}) is written as 
\begin{equation}
	x_{qs}(k+1)=x_{qs}(k)+hBu(k).
	\end{equation}
Taking the $81$ activation patterns $\{-1,0,1\}^4$ as inputs, we see that at each time step, (\ref{equation:qs}) can move in any of the $25$ directions (including a zero vector) depicted in Fig. \ref{fig:dir}.
    	\end{example}
    
    Consider solutions to the linear ordinary differential equation (ODE) (\ref{equation:LTI}). The goal of the general emulation problem is to find piecewise constant quantized inputs with sampling interval $h>0$ such that the resulting trajectories of (\ref{equation:qs}) with initial state $x_{qs}(0)=x_0$ approximate the continuous system $(\ref{equation:LTI})$. Here, we solve the emulation problem that finds a partition of the state space 
    $\{U_i\,:\, \cup\  U_i = \mathbb{R}^n;\ \ U_i^o\cap U_j^o=\emptyset;\ U_i^o={\rm interior}\ U_i\}$
    and a selection rule depending on the current state $x_{qs}(k)$ for assigning values of the input at the $k$-th time step to be $u(k)\in{\cal U}=\{-1,0,1\}^m$, so that for each $x_{qs}\in U_i$, $e^{Ah}x_{qs}(k)+\int_{0}^{h}e^{A(h-s)}dsBu(k)$ is as close as possible to the emulated LTI system. We have defined various metrics in terms of which we determine the fitness of an approximation such as the direction and magnitude between two systems as was studied in \cite{baillieul2021neuromimetic}. When $m$ is large ($B$ is $n \times m$ with $m \gg n$), it will frequently be the case that several activation patterns have comparable fitness. Because two or more activation patterns may give approximately equal quality of emulation, it may be the case that over the course of a trajectory, we need to consider
    multi-step fitness. This is illustrated in the Fig.\ref{fig:dir}, where we see that the red approximants have the best first vector step matching the vectorfield (black vector), but in considering pairs of steps, the blue pair of vectors end up closer to the black pair than the red. Hence, when applying algorithms to learn to optimally emulate a given system, we need to focus on methods that consider multiple steps, and we need to be aware that there may be multiple solutions of approximately equal value. 
   
	 Our goal is to design algorithms that can meet the following objectives:
    \begin{itemize}
    	\item Quantized system (\ref{equation:qs}) can emulate the LTI system (\ref{equation:LTI}), i.e., at any given point, the model generated by the algorithm can compute an optimal quantized direction taking into account multiple steps.
    	\item The emulation task can still be achieved when there are channel dropouts in the quantized system during the emulating process.
    	\item When the emulated LTI system has changed, the model still works well under new circumstances directly without the need of retraining.

	\end{itemize}

	\section{ Learning methods for Emulation Problems}
	For emulation learning of trajectories of a given LTI system, we propose two approaches: an MPC data-based supervised learning approach that trains a neural network using training data collected from solving the MPC optimization problem and a generalized DQN algorithm by exploring the environment. We shall discuss the advantages of these approaches in terms of the optimality of trajectory, computational efficiency and resilience to channel dropouts.
	%We propose a generalized DQN-like Algorithm to achieve the whole learning process, which can meet these objectives.
	\subsection{MPC Data-based Supervised Learning}
	As described in \cite{Sun2022MPC}, the Model Predictive Control (MPC) approach can be formulated as solving the following optimization problem: 
	\begin{eqnarray}
		\label{MPC_optimal}
			\min\limits_{u_{0|k},...,u_{N-1|k} }J&=&|u_{n|k}|^2_R+|x_{N|k} -x_{ref}(N|k)|^2_P\nonumber\\
			&&\ \ \ \ \ +\sum_{n=0}^{N-1}|x_{n|k}-x_{ref}(n|k)|^2_Q\nonumber\	\\
			s.t. &\quad& u_{n|k}\in\left\{-1,\ 0,\ 1\right\}^m,\nonumber\\
			&\quad& x_{n+1|k}=Ax_{n|k}+Bu_{n|k},\nonumber\\
			&\quad& x_{ref}(n+1|k)=e^{Hh}x_{ref}(n|k),\nonumber\\
			&\quad& \quad \quad \quad \quad\forall n= {0,1,...,N-1},
	\end{eqnarray}
	where $P,Q, R$ are positive definite matrices and the function $|x|^2_P=x^TPx$, with similar definitions for $Q$ and $R$. The reference system $x_{ref}$  is the linear time-invariant system (\ref{equation:LTI}).
	
	By solving the optimization problem (\ref{MPC_optimal}), we can obtain the optimal direction for the quantized system to track. However, this problem is integer programming, which is a computational challenge, especially since the dimension $m$ is large in our case. 
	
	We approach the problem by means of a neural network with training data comprised of many solutions that have been obtained off-line. The idea is that at run time the trained model will directly select appropriate activation patterns at each time step, without the need to resolve (\ref{MPC_optimal}). Once the model is trained, it can steer (\ref{equation:qs}) to emulate (\ref{equation:LTI}) in a way that is optimal or nearly optimal in terms of (\ref{MPC_optimal}). Unfortunately, collecting sufficient training samples remains prohibitive. Nevertheless, the kinds of resilience encountered in previously studied overcomplete systems, \cite{baillieul2021neuromimetic}, \cite{Sun2022} are seen to be present. From the simulation experiments in Section V below, we find that even when there were channel dropouts (simulating neuronal damage), the systems were able to choose new quantized trajectories emulating the the desired asymptotically stable motion.
	
	In an attempt to find an approach that is both more computationally tractable and more similar to learning mechanisms of neurobiology, we next propose a deep Q-learning method that works well in the presence of channel dropouts at both the training and final execution stages.
	
	%Therefore, we construct a neural network to tackle this problem by simultaneously collecting data points from solving the optimal MPC problem. Then the trained model can choose the activation pattern directly at each step based on the current metric -- the objective function of (\ref{MPC_optimal}) without solving the MPC problem. The main advantage of this method is that even at the beginning of emulation, the quantized system obtains the optimal quantized direction from the hard solution of MPC instead of interacting with the environment and receiving feedback. After collecting enough data points, it can use the output of the neural network to reduce the running time. However, the data-collecting part is still prohibitive. Meanwhile, from the simulation in Section V, we found that the structure of the neural network needs to be complicated to guarantee the loss is small. From the experiments, we observed that the animals showed a high degree of cooperation and efficiency in selecting the pathway when completing a complex sequence of actions. Even in states with neuronal damage, they were able to immediately find other pathways to complete the task. This approach does not explain well the intrinsic mechanism of neurobiology. Therefore, we propose another deep reinforcement learning method that can work well even if channels are intermittently unavailable, which exhibit the learning and relearning feature.
	
	\subsection{Generalized DQN-like Algorithm}
	Reinforcement learning using deep neural networks has been proven successful from experiments in various areas recently \cite{Levine2018}. The basic idea of RL is that an agent and an environment interact continuously. The agent receives a state ($s_t$) at step $t$ from the environment and chooses an action ($a_t$), then the environment reacts to this action leading to a new state ($s_{t+1}$) for the agent along with a reward ($r_t$) that reflects the value of the action taken. Q-learning is one of the most efficient strategies for carrying out this type of learning \cite{john1989}.  It operates using a $Q$ function that stores states-action pairs and maps them to Q values defined by
	 	\begin{equation}
	 		\label{Q_function}
	Q_\pi(s_t,a_t)=\mathbb{E}[r_t+\sum_{i=1}^{T-t}\gamma^ir_{t+i}],
\end{equation}where $\gamma<1$ is a discount factor. Since in in what follows we only consider the deterministic case, the notation $\mathbb{E}[\cdot]$ can be ignored. The goal is to find an optimal policy $\pi^*$ that can maximize the $Q_\pi$. The optimal $Q$ function satisfies the Bellman's equation:
	\begin{equation}		
	Q^*(s_t,a_t)=r_t+\gamma\max\limits_{a'} Q^*(s_{t+1},a'). 
\end{equation}
An iterative update of the $Q$ function is given by
\begin{equation}		
	\label{loss}
	loss=(r_t+\gamma \max\limits_{a'}\ Q(s_{t+1},a')- Q(s_t,a_t))^2. 
\end{equation}
Iterating updates using this state-action pair expression is only practical when the number of these pairs is finite and not too large. When the state and action space have large numbers of elements or perhaps take on continuous values, it becomes infeasible to construct a $Q$ table. Therefore, a neural network with learnable parameters $\phi$ is used to approximate the complex nonlinear $Q$ function. The learning rule for $\phi$ with learning rate $\alpha$ is constructed as
\begin{equation}
	\label{learning_rule}
	 \phi_{i+1}=\phi_{i}-\alpha\nabla_{\phi_i} (r_t+\gamma \max\limits_{a'}\ Q_{\phi_i}(s_{t+1},a')- Q_{\phi_i}(s_t,a_t))^2.
	\end{equation}
It is well-known that direct iteration of (\ref{learning_rule}) may not converge since the term $\max\limits_{a'}\ Q_{\phi_i}(s_{t+1},a')$ also depends on $\phi_i$, \cite{cDQN}. The deep Q network (DQN) algorithm solves this problem by introducing another neural network called the {\em target network} with parameters $\hat \phi$ to predict the target Q value. It can stabilize the learning by replacing the term $\max\limits_{a'} Q_\phi(s_{t+1},a')$ with $\max\limits_{a'} Q_{\hat \phi}(s_{t+1},a')$. Therefore, when learning parameters $\phi$, only the part $Q_{\phi}(s_t,a_t)$ changes, keeping the target function $(r_t+\gamma \max\limits_{a'}\ Q_{\hat\phi}(s_{t+1},a')$ fixed to avoid oscillations. Therefore, the loss function of the DQN algorithm is
\begin{equation}
	\label{l_DQN}
	l_{DQN}(\phi;\hat\phi)=(r_t+\gamma \max\limits_{a'}\ Q_{\hat\phi}(s_{t+1},a')- Q_{\phi}(s_t,a_t))^2.
\end{equation}
For every $C$ steps, the parameters $\phi$ from the prediction network are copied to the target network $Q_{\hat \phi}$.

The efficiency of the DQN algorithm has been illustrated in several experiments, such as playing Atari games \cite{DQN} and planning vehicle routing \cite{vehicle}. Inspired by these successes, we proposed a DQN-like algorithm to solve the restricted emulation problem in \cite{Sun2022}, which focused on one-step optimization of quantized system motion starting from points on the unit sphere. In what follows, we provide an extension by proposing a generalized DQN-like Algorithm \ref{alg:general_DQN} for the quantized system to learn the whole emulation trajectory for any linear dynamic system. To guarantee this emulation problem is Markov Decision Process (MDP) \cite{bellman1957}, the state $s_t$ needs to be designed carefully, where it has information of two systems and the time. Therefore, $s_t\in\mathbb{R}^{2n}$ in this paper contains two parts: the error vector between two systems, i.e., $x(t)-x_{qs}(t)$ and the state of the emulated LTI system $x(t)$. The action space is a set of quantized directions ($Dir$) so that we use $d_t$ (as illustrated in Fig. \ref{fig:dir}) instead of $a_t$. Therefore, in each transition, when $s_t$ and the action $d_t$ are given, we can determine the next state $s_{t+1}$. The reward $r_t$ is a function involving the $L_2$ norm of the error, which is also determined. The quantized direction is obtained by following an $\epsilon$-greedy policy, which is 
\begin{equation}
	\label{policy}
	d_t=\begin{cases}
		\text{choose}\ d\in Dir\ \text{randomly},\quad p=\epsilon \\
		\arg\max\limits_{d}Q^*_{\phi}(s_t,d),\quad p=1-\epsilon
	\end{cases}
	\end{equation}
 Since the convergence of this DQN algorithm has only been demonstated by experiments and is known sometimes diverge \cite{Tsitsiklis1997}, \cite{Van2018}, a new loss function from \cite{cDQN} is utilized:
 \begin{equation} 
 	\label{loss_f}
 	Loss(\phi;\hat\phi)=\max\{l_{DQN}(\phi;\hat\phi),l_{MSBE}(\phi)\},
 \end{equation}
 where 
\begin{equation} 
	\label{l_MSBE}
	l_{MSBE}(\phi)=(r_t+\gamma \max\limits_{d'}\ Q_{\phi}(s_{t+1},d')\\- Q_{\phi}(s_t,d_t))^2
\end{equation}
	denotes the mean squared Bellman error. By using this loss function (\ref{loss_f}), our Algorithm \ref{alg:general_DQN} is guranteed to converge since $\hat\phi_{i+1}=\arg\min\limits_{\phi}Loss(\phi;\hat\phi_{i})$ and
	\begin{equation}
		\begin{split}
			\min \limits_{\phi}Loss(\phi;\hat\phi_{i+1})&\le Loss(\hat\phi_{i+1};\hat\phi_{i+1})= l_{MSBE}(\hat\phi_{i+1})\\
			&\le Loss(\hat\phi_{i+1};\hat\phi_{i})=\min \limits_{\phi}Loss(\phi;\hat\phi_{i}).
			\end{split}
	\end{equation}
	
	\begin{algorithm}
		\caption	{Learning optimal path to emulate dynamic systems}\label{alg:general_DQN}
		\begin{algorithmic}[1]
			\State Input: Activation patterns $U=\{u_1,u_2,...,u_K\}$ and its direction vectors of quantization output alphabet to form an action space $Dir=\{d_1,d_2,...,d_{K'}\}$, a learning metric $G$;
			\State Initialize replay memory $D$ with capacity $N$;
			\State Initialize action-value $Q$ function with parameter $\phi$  and target-$Q$ function with parameter $\hat\phi=\phi $;
			%\While{True}
			
			\For{\texttt{episode= 1,M}}
			\State Start from the initial state $s_0=\{e(0),x_{ref}(0)\}$, which contains the initial error vector between emulated and quantized systems and the location of the reference system; 
			\For{\texttt{t= 1,T}}
			\State Observe two systems and record the error vector and location of emulated system as$s_t=\{e(t),x_{ref}(t)\}$. Choose direction $d_t=\pi^\epsilon(s_t)$, get reward $r_t=-G(s_t)$ and new state $s_{t+1}=\{e(t+1),x_{ref}(t+1)\}$;
			\State Store $(s_t, d_t, r_t,s_{t+1})$ as a memory cue to $D$;
			\State Sample random minibatch of cues  $(s_j, d_j, r_j,s_{j+1}),j\in[0,t]$ from $D$;
			%\State $y_{\hat\phi}=	r(j)+\gamma \max\limits_{d'}\ Q_{\hat\phi}(s(j+1),d')$;
			%\State $y_{\phi}=	r(j)+\gamma \max\limits_{d'}\ Q_{\phi}(s(j+1),d')$;
			%\State $l_{DQN}(\phi;\hat\phi)=(y_{\hat\phi}- Q_{\phi}(s(j),d(j)))^2$;
			%\State $l_{MSBE}(\phi)=(y_{\phi}- Q_{\phi}(s(j),d(j)))^2$;
			\State Apply the loss function $Loss(\phi;\hat\phi)=\max\{l_{DQN}(\phi;\hat\phi),l_{MSBE}(\phi)\}$ to train $Q$ network and every $C$ steps,  $\hat\phi\leftarrow\phi$;
			%$\hat\phi=\arg \min\limits_\phi Loss(\phi;\hat\phi)$;
			\EndFor
			\EndFor
		\end{algorithmic}
	\end{algorithm}	
	%In the algorithm, $l_{MSBE}=(r(t)+\gamma \max\limits_{d'}\ Q_{\phi}(s(t+1),d')- Q_{\phi}(s(t),d(t)))^2$ denotes the mean squared Bellman error, and $l_{DQN}$ is the loss function of the DQN algorithm. By choosing the loss function in our emulation algorithm to be $Loss(\phi;\hat\phi)=\mathbb E[\max\{l_{DQN}(\phi;\hat\phi),l_{MSBE}(\phi)\}]$, the algorithm is guaranteed to converge, which has been proved in \cite{cDQN}.
	
	\begin{remark}
		It is noted that Algorithm \ref{alg:general_DQN} can not only be used for the quantized system to learn the trajectory of LTI systems but also any given system. However, since the quantized direction set is finite, when emulating unstable or nonlinear systems, the emulation performance can only be guaranteed in a local sense. An example of using this algorithm to track a nonlinear system around equilibria will be considered elsewhere.
		\end{remark}

 \begin{remark} ({\em On resilient learning})
	When using our deep Q network to learn the trajectories, channels can drop out at any time without causing problems. Since at each step, it chooses the optimal available action (direction) by ($\ref{policy}$). When there are channels dropout and the optimal quantized direction is unavailable, the policy can generate the direction with the largest $Q$ value from all available candidates. %Meanwhile, this algorithm is adaptive to the situation of changing time-step. When the system approaches the origin, shorter quantized directions are needed. An approach is decreasing the time step from $h$ to $h'<h$ to decrease the magnitude of quantized directions. Instead of choosing the action $d(t)=\pi^\epsilon(x(t))$, we choose the direction $d(t)=\frac{h'}h\pi^\epsilon(x(t)\frac{h}{h'})$.
	\end{remark}

\section{Mapping-based Transfer Learning}
In addition to the ability to learn, neurobiological system have the ability to generalize and adapt what they have learned to new but similar problem domains. To explore such {\em transfer learning} in the context of our emulation problems, suppose that a DQN has been trained for emulating a particular LTI system. Suppose another LTI system has the  form
\begin{equation}
	\label{equation:new_LTI}
	\dot z=H_o z,
\end{equation}
and $H_o=OHO^{-1}$ where $O$ is known and invertible. To emulate this LTI system, we utilize a mapping-based transfer learning method \cite{tan2018}. The coordinate is transformed by setting $z=Ox$, then the dynamic equations of these two LTI systems have the following relation: 
\begin{equation}
	\label{equation:transfer}
	\begin{aligned}
		&\dot z=H_oz=OHO^{-1}z=OHO^{-1}Ox=OHx,\\
		&\dot z=O\dot x=OHx
	\end{aligned}
\end{equation}
	Assume we have already obtained a trained model from the MPC data-based or the DQN-like algorithm for the emulation problem of (\ref{equation:LTI}) using (\ref{equation:qs}), say $F:\mathbb{R}^{2n}\rightarrow\mathbb{R}^{n}$, which is expressed by a neural network. $2n$ is the dimension of state $s$ and $n$ is the dimension of the quantized direction $d$. The model can predict the optimal quantized direction at any given state $s$. For example, $d=\arg\max\limits_{d}Q^*_{\phi}(s,d)$ if the DQN model is used.
	%For example, if the neural network has $k$ linear layers and ReLU activation functions between two layers The output dimension equals the total number of elements in the action space and the output value denotes the fitness value for each action. Then $F(s_t)=\arg \max[ w_kReLU(w_{n-1}ReLU(\cdots ReLU(w_1s_t+b_1))+b_{k-1})+b_k]$,where $\{w_i,b_i\},i=1,2,\cdots k$ are learnable parameters. We try to obtain a selecting policy $F_o$ from $F$.

	Instead of learning $F_o$ by constructing another dataset by solving a new MPC problem or learning a new deep Q network---which are time-consuming, we try to obtain a selection policy $F_o$ from $F$. The strategy will be to use mapping-based transfer learning, \cite{Hessel2018}. The main idea is to express the relationship of features. Therefore, $F$ can be used directly by the following steps to obtain the learned policy $F_o$:
	\begin{itemize}
		\item Record the learning metric like the error vector between the new LTI system (\ref{equation:new_LTI}) and the quantized system (\ref{equation:qs}) as feature $f_1\in\mathbb{R}^n$, the direction (i.e., $[x(k+1)-x(k)]/h$) or the location of the LTI system (i.e., $x(k)$) as feature $f_2\in\mathbb{R}^n$. Combine these features to be $f_o=[f_1;f_2]\in\mathbb{R}^{2n}$;
		\item Transform the coordinate of features $f_o$ to the coordinate system of (\ref{equation:LTI}): $f=\begin{bmatrix}
 		O^{-1} & 0\\
 		0&O^{-1}
 	\end{bmatrix}{f_o}$;
		\item Predict the optimal direction by $\vec{d}=F(f)$;
		\item Change back the predicted direction to the coordinate system of (\ref{equation:new_LTI}): $\vec{d_o}=O\vec{d}$. It is noted that $\vec{d_o}$ may not be in the quantized direction set formed in the current coordinate;
		\item Find the nearest neighbor of $\vec{d_o}$ in the direction set by kd-tree or Hebbian-Oja algorithm \cite{Sun2022}, and denote this by $\vec{d^*}$.
        \item It is the (sub)optimal output direction for emulating the system (\ref{equation:new_LTI}).
\end{itemize}

\begin{figure}[h]
	\begin{center}
		\includegraphics[scale=0.7]{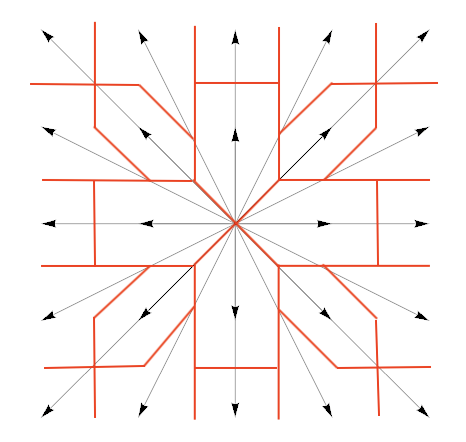}
	\end{center}
	\caption{An example of a mapping function that partitions the space according to the closest quantized direction (not including zero direction). The black vectors are the quantized directions formed by $Bu$ corresponding to Fig. 1(a), and the red lines are the division boundaries between the cells $U_i$. }
	\label{fig:neighbor}
\end{figure}

\begin{lemma}
	\label{transfer1}
	Following the above steps, the learned policy that gives $\vec{d_o}$ is $\tilde{F_o}=OF(\begin{bmatrix}
 		O^{-1} & 0\\
 		0&O^{-1}
 	\end{bmatrix}f_o)$. If we already have a mapping rule $M$, as shown in Fig. \ref{fig:neighbor}, which computes the nearest available quantized direction, we can write the learned policy for the new emulation problem to be $F_o=M(OF(\begin{bmatrix}
 		O^{-1} & 0\\
 		0&O^{-1}
 	\end{bmatrix}f_o))$.
\end{lemma}
\begin{proof}
By applying the transformation steps on the given model, the formula for $F_o$ is easily derived, and the details are omitted.
	\end{proof}

It can be observed that there is a special case when the transform matrix $O$ makes no change of the action space, which is the set of all combinations of $Bu$. In this case, $Dir$, is the same as the set $Dir_o$ formed by $OBu$. Therefore, $\vec{d_o}=\vec{d^*}$ and the last step to find the nearest neighbor is no longer needed. Section V provides a simulation of such a case.

Let $\pi^*$, $Q_\phi^*$ denote the optimal policy and optimal $Q$ function for the original emulation problem, and $\pi^*_o$, $Q_{\phi_o}^*$ for the new one. According to the definition,
\begin{equation}
	\label{eq:pi}
	\begin{split}
	\pi^*(s)&=\arg\max\limits_{d}Q_\phi^*(s,d),\\
	 \pi_o^*(s_o)&=\arg\max\limits_{d_o}Q_{\phi_o}^*(s_o,d_o).
	 \end{split}
	\end{equation}
\begin{theorem}
	\label{theorem}
	 %The $Q-$network for this new special case problem can be expressed as $Q^*_{\phi_o}(s_o,d)=Q^*_\phi(O^{-1}s_o,d)$ and hence, with the convergence of $Q^*_\phi$ for the original problem, $Q-$network for the new emulation problem also converges.
	 The optimal learning policy for this new emulation problem can be expressed as $\pi_o^*(s_o)=O\pi^*(\begin{bmatrix}
	 	O^{-1} & 0\\
	 	0&O^{-1}
	 \end{bmatrix}s_o)$, when the following conditions are satisfied:\\
 	(a) The direction space is invariant before and after the transformation.\\
 	(b) The first layer of the $Q_\phi$ network is linear and $Q_{\phi_o}$ has the same structure as $Q_\phi$.\\
 	(c) The reward function for the new problem is designed to be $r(\begin{bmatrix}
 		O^{-1} & 0\\
 		0&O^{-1}
 	\end{bmatrix}s_o)$, where $r(\cdot)$ is the reward function of original problem.\\
\end{theorem}
\begin{proof}
	From the basic setup of Q-learning, we have
	\begin{equation} 
		\begin{split}
			\nonumber
		Q_\phi^*(s,d)&\quad=\quad r(s)+\gamma\max\limits_{d'}Q_\phi^*(s',d'),\\
		\end{split}
		\end{equation}
		 where $s'$ is the state after the system at state $s$ taking direction $d$. Here we only consider deterministic policy so that $s'$ is also deterministic.  For the new problem, we first transfer the coordinate to the original one. Since state $s\in\mathbb{R}^{2n}$ contains the error vector as well as the current location of the LTI system, $s_o\rightarrow\begin{bmatrix}
		 	O^{-1} & 0\\
		 	0&O^{-1}
		 	\end{bmatrix}s_o$ and $d_o\rightarrow O^{-1}d_o$. Then,
	\begin{equation}
		\label{eq:Q}
		\begin{split}
			Q_\phi^*(\begin{bmatrix}
				O^{-1} & 0\\
				0&O^{-1}
			\end{bmatrix}&s_o,O^{-1}d_o)= r(\begin{bmatrix}
			O^{-1} & 0\\
			0&O^{-1}
		\end{bmatrix}s_o)\\
			&+\gamma\max\limits_{O^{-1}d'}Q_\phi^*(\begin{bmatrix}
				O^{-1} & 0\\
				0&O^{-1}
			\end{bmatrix}s'_o,O^{-1}d').
		\end{split}	
	\end{equation}
	If the first layer of $Q_\phi$ is linear, say $\{\omega_{1i} s+b_{1i}\}$, $\forall i=1,2,\cdots k$, then $\omega_{1i}$ can absorb the transformation matrix and becomes $\omega_{1i}\begin{bmatrix}
		O^{-1} & 0\\
		0&O^{-1}
	\end{bmatrix}$, where $k$ is the number of units in the first layer.
	 Assume all other parameters are the same as in $Q_\phi$ except {$\omega_{1i}$}.  We denote this new family of parameters to be $\phi_n$. Fig. \ref{fig:Q} shows the detail of such parameters transformation. 
	 Then, ($\ref{eq:Q}$) becomes
	\begin{equation}
		\label{eq:Q_n}
		\begin{split}
			Q_{\phi_n}^*(s_o,O^{-1}d_o)\quad&=\quad r(\begin{bmatrix}
				O^{-1} & 0\\
				0&O^{-1}
			\end{bmatrix}s_o)\\
			&\quad\quad+\gamma\max\limits_{O^{-1}d'}Q_{\phi_n}^*(s'_o,O^{-1}d').
		\end{split}	
	\end{equation}
%\begin{comment}
	\begin{figure}[h]
		\begin{center}
			\includegraphics[scale=0.45]{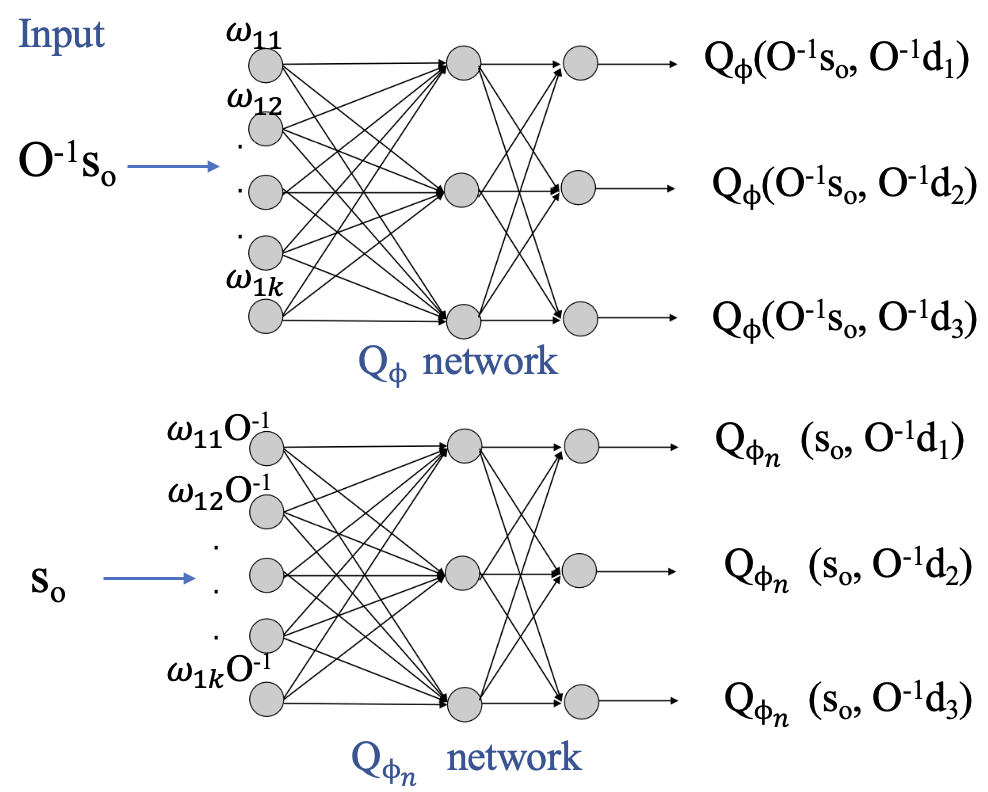}
		\end{center}
		\caption{The parameters tranformation from $Q_\phi$ to $Q_{\phi_n}$ .}
		\label{fig:Q}
	\end{figure}
%\end{comment}
	
 Since the direction space is invariant under matrix $O$, $O^{-1}d_o$ and $d$ have the same space, we can also write (\ref{eq:Q_n}) as
		\begin{equation}
		\begin{split}
			Q_{\phi_n}^*(s_o,d)=r(\begin{bmatrix}
				O^{-1} & 0\\
				0&O^{-1}
			\end{bmatrix}s_o)+\gamma\max\limits_{d'}Q_{\phi_n}^*(\tilde s_o,d'),
		\end{split}	
	\end{equation}
where $\tilde s_o$ is the state after the new system at state $s_o$ taking direction $d$.
 From the uniqueness of the optimal Q-network, it is proved that $Q^*_{\phi_o}(s_o,d_o)$ = $Q^*_{\phi_n}(s_o,d_o)$ as long as the reward function in the new problem is $r(\begin{bmatrix}
	 	O^{-1} & 0\\
	 	0&O^{-1}
	 \end{bmatrix}s_o$).
 
  From (\ref{eq:pi}) and the invariant direction space, we have 
 \begin{equation}
 	\begin{split}
 \pi_o^*(s_o)&=\arg\max\limits_{d_o}Q_{\phi_o}^*(s_o,d_o)\\
 &=O\arg\max\limits_{O^{-1}d_o}Q_{\phi_n}^*(\begin{bmatrix}
 	O & 0\\
 	0&O
 \end{bmatrix}s,O^{-1}d_o)\\
&=O\arg\max\limits_{d}Q_{\phi}^*(s,d)=O\pi^*(s)\\
&=O\pi^*(\begin{bmatrix}
	O^{-1} & 0\\
	0&O^{-1}
\end{bmatrix}s_o)
 \end{split}
 \end{equation}
\end{proof}

\begin{lemma}
	$Q_{\phi_o}^*$ is convergent.
	\end{lemma}
\begin{proof}
	From the proof of Theorem $\ref{theorem}$, we obtain the relationship between $\phi_o$ and $\phi$ when both of them construct the optimal $Q$ function. Therefore, from the convergence of $Q^*_\phi$, $Q_{\phi_o}^*$ is also convergent.
	\end{proof}
For general cases, in the last step above, finding the nearest neighbor of $\vec{d_o}$ is a challenging task when there are numerous candidate quantized directions if we use the exhaustive search. We may use the Hebb-Oja algorithm which has been introduced in \cite{Sun2022,Oja1982} to compute the nearest direction through iterations. Another approach is construsting a kd-tree, which is a binary space partitioning data structure. Each non-leaf node can be thought of as a dividing hyperplane. The points on one side of this hyperplane are represented by the left subtree of the node, while the right subtree represents the points on the other side of the hyperplane. Each node in the tree is associated with one of the k dimensions, and the hyperplane is perpendicular to the axis of that dimension. Here, the quantized directions can be viewed as nodes and the direction $\vec{d_o}$ is the search key. This tree structure is well suited for channel dropout situations because when nodes in the tree have been removed, instead of destroying the whole structure, we only need to form the set of all nodes and leaves from the children of the removed nodes and recreate that part of the tree. 

By applying either of two approaches mentioned above, the mapping function $M$ is obtained. We can directly use this model in Lemma \ref{transfer1} to compute a (sub)optimal direction. In addition, it can also be used to learn the Q-network by the same structure of loss function (\ref{loss_f}) and the initial $\phi_o=\phi$. From experients in Section V, it can be observed that with these initial parameters, the training efficiency improves compared with a randomly generated one.

\section{Simulation and Analysis}
\begin{figure}[h]
	\begin{center}
		\subfigure[]{
		\includegraphics[height=0.2\textwidth]{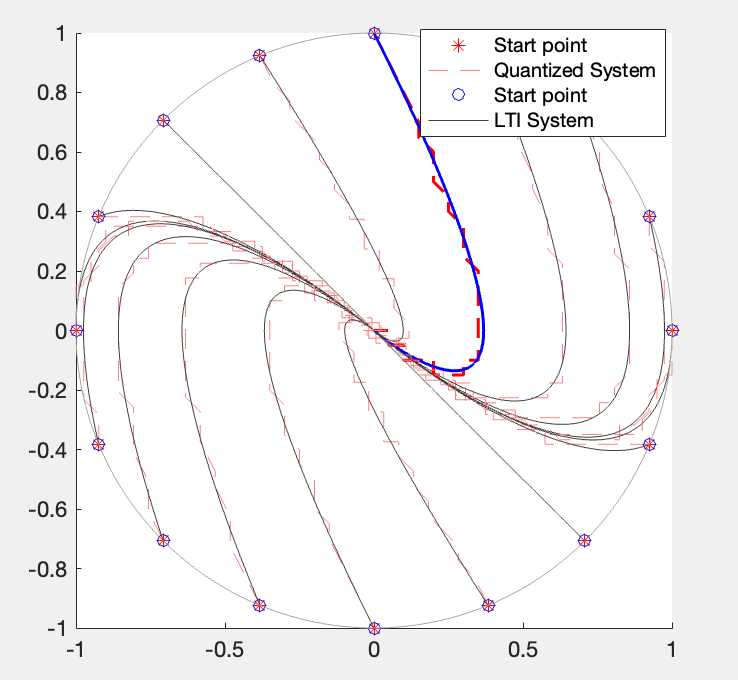}}
	\subfigure[]{
		\includegraphics[height=0.2\textwidth]{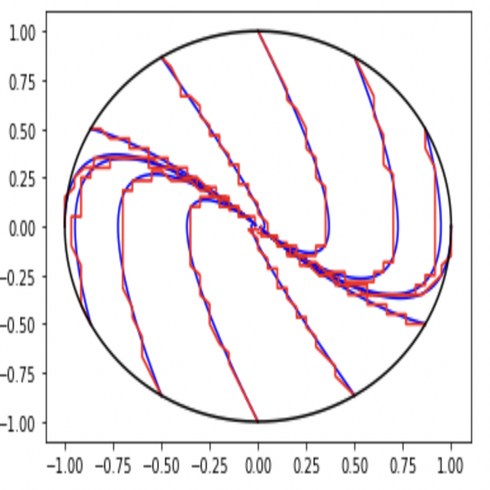}}
	\end{center}
	\caption{Figure (a) is the emulation result when solving the integer optimization problem ($\ref{MPC_optimal}$) directly. Blue trajectories are the LTI system and red ones are the quantized system. Figure (b) is the emulation result using the MPC data-based learned model.}
	\label{fig:MPC}
\end{figure}

In this section, we provide simple simulations of the learned trajectory per the MPC-based method and per the DQN algorithm. The LTI system we try to emulate is $\dot x=Hx=\begin{bmatrix}
	0& 1\\
	-1&-2
\end{bmatrix}x$ with all its eigenvalues located in the left-half plane. To simplify the example, we set $A$ and $B$ be as in Example 1 %to be zero in (\ref{equation:qs}), $B=\begin{bmatrix}
	%1& 0 & -1 & 0\\
	%0&  1&  0&-1
%\end{bmatrix}$ 
and choose time step $h=0.05$. In the cost function, we design $P=Q=\begin{bmatrix}
	5 &0 \\
	0&5
\end{bmatrix}$ and $R$ to be $0.05*\mathbb{I}_m$. The result is shown in Fig. \ref{fig:MPC}(a). It can be observed that by solving the MPC optimization problem (\ref{MPC_optimal}), we generate a sequence of vector steps that emulate the LTI system moving towards the origin.

At the same time, we collect a total of 10000 data points with error vector, i.e., $x(k)-x_{qs}(k)$ and directions, i.e., $[x(k+1)-x(k)]/h$ and $[x_{qs}(k+1)-x_{qs}(k)]/h$ of two systems as features, and quantized directions as their labels. Then, we construct a regular densely-connected four-layer neural network with ReLU, Sigmoid, or Linear as their activation functions. The number of nodes in each layer is 1200, 1200, 1200, and 25, respectively. After 20 training epochs, we obtain a model with a training accuracy of 94.1\%. The test dataset comes from another emulation with all initial points in the unit circle, which contains 840 data points, and the accuracy is 90.7\%. The trajectory generated by this model is shown in Fig. \ref{fig:MPC}(b). 

\begin{figure}[htbp]
	\centering
		\subfigure[]{
		\includegraphics[height=0.2\textwidth]{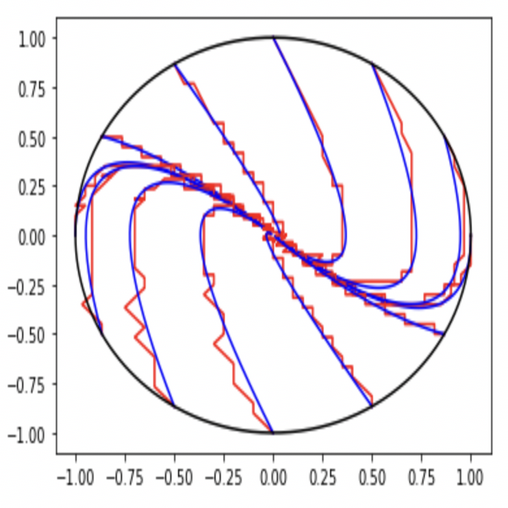}}
	\subfigure[]{
		\includegraphics[height=0.2\textwidth]{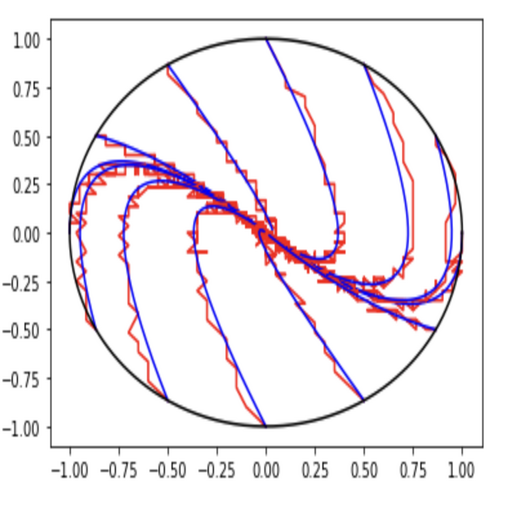}}
	\caption{Figure (a) is generated by the DQN algoirhm with time step $h=0.05$, while figure (b) is using the same model but having channel dropouts.}
	\label{fig:model}
\end{figure}

Using the same $A,B$ matrices and the same emulated system, we also construct a two linear-layer $Q$ network with a hidden layer with 200 activation units between them to learn the trajectory. The activation function is ReLU. Fig. \ref{fig:model}(a) is an emulating system produced by the generalized DQN algorithm. The learning metric in the algorithm only considers the $L_2$ norm of the error vector between two systems. At time $T$, the location of the LTI system is $x(T)=e^{THh}x_0$, while the quantized system is in the location $x_{qs}(T)=x_0+hBu(0)+hBu(1)+\cdots+hBu(T-1)$. Therefore, $G(u,T)=||e^{THh}x_0-(x_0+hB\sum_{t=0}^{T-1}u(t))||^2$, where $x_0$ is the initial point. In this simple example, there are 25 distinct directions (including zero vector) in the quantization output alphabet formed by $Bu$, where $u_i\in\{-1,0,1\}^4$. Fig. \ref{fig:model}(b) provides an example of having one channel randomly dropout at each time instant. Though it has degraded performance, it still has the ability to emulate.

\begin{figure}[htbp]
	\centering
	\subfigure[]{
		\includegraphics[height=0.2\textwidth]{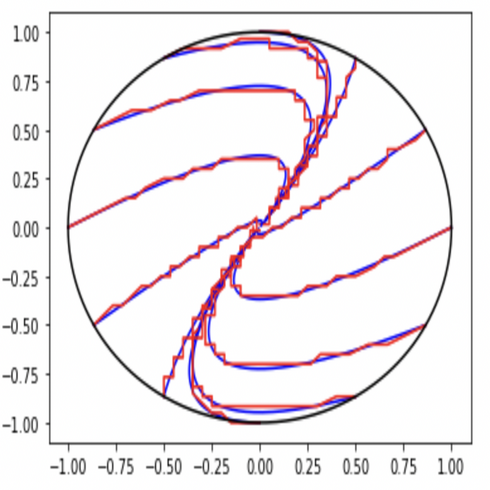}}
	\subfigure[]{
		\includegraphics[height=0.2\textwidth]{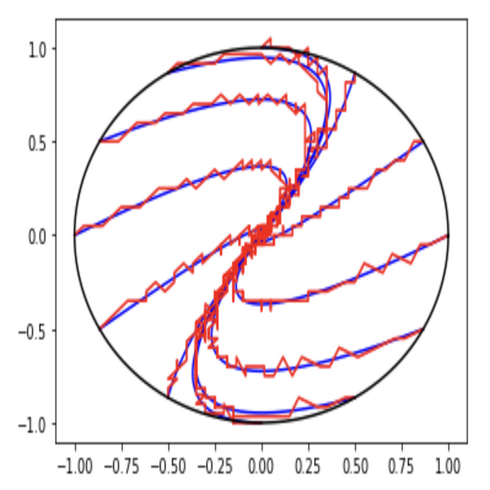}}
	\caption{Figures show the emulation trajectory of another LTI system $\dot z=H_oz=\begin{bmatrix}
			-2& 1\\
			-1&0
		\end{bmatrix}z$ by transferring the MPC-based and DQN model.}
	\label{fig:transfer}
\end{figure}

Next, we use the mapping-based transfer learning method to emulate another LTI system $\dot z=H_oz=\begin{bmatrix}
	-2& 1\\
	-1&0
\end{bmatrix}z$. The transform matrix is $O=\begin{bmatrix}
0& 1\\
-1&0
\end{bmatrix}$ in (\ref{equation:transfer}), which results in the action space being invariant. The models we use are the MPC data-based learning and the DQN-like algorithm, respectively. The tracking performance is shown in Fig. \ref{fig:transfer}. It illustrates that both of these two models can be transferred to solve another emulation problem.

\begin{figure}[htbp]
\centering
\subfigure[]{
		\includegraphics[height=0.2\textwidth]{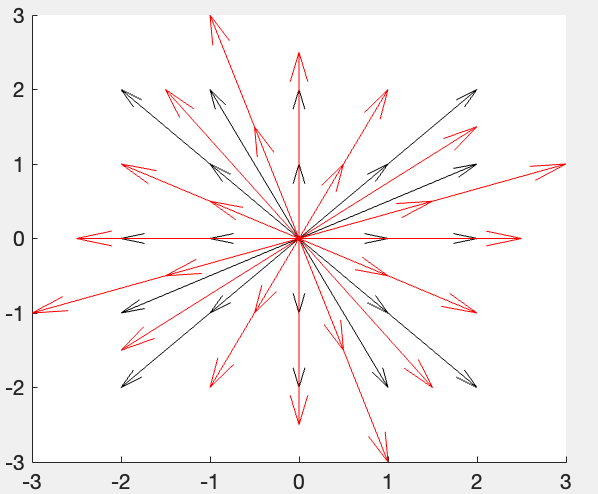}}
	\subfigure[]{
		\includegraphics[height=0.2\textwidth]{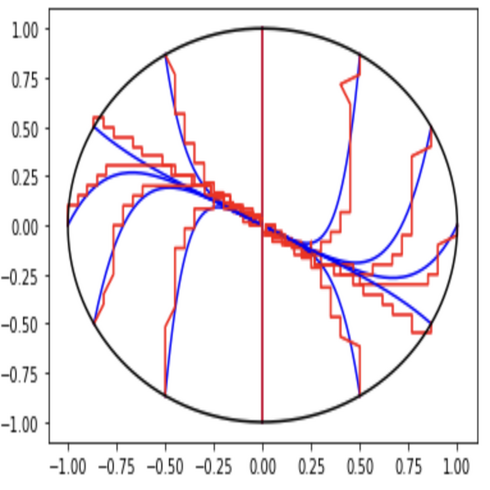}}

	\caption{Figure (a) shows the new sequence of quantzied directions in red to emulate the LTI system $\dot z=H_oz=\begin{bmatrix}
			-0.5& 0\\
			-1&-2.5
		\end{bmatrix}z$, while black ones are the previous directions to be used to train the DQN model. Figure (b) is the emulation trajectory using new vector directions.}
	\label{fig:transfer2}
\end{figure}

When the transform matrix is $O=\begin{bmatrix}
	1& 0.5\\
	-0.5&1
\end{bmatrix}$ in (\ref{equation:transfer}), the action space is no longer invariant as is illustrated in Fig. \ref{fig:transfer2}(a). The red vector directions form the new action space, used in emulating a new LTI system. For this example, the new emulated LTI system is $\dot z=H_oz=\begin{bmatrix}
-0.5& 0\\
-1&-2.5
\end{bmatrix}z$. The model we used is the DQN-like Algorithm \ref{alg:general_DQN} and the tracking performance is shown in Fig. \ref{fig:transfer2}(b).

\begin{figure}[htbp]
	\centering
		\subfigure[]{
		\includegraphics[height=0.2\textwidth]{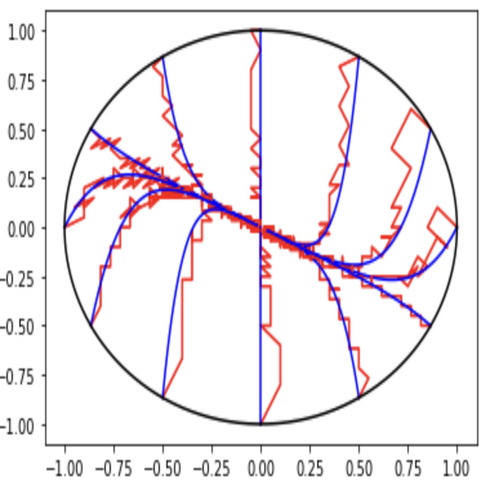}}
	\subfigure[]{
		\includegraphics[height=0.2\textwidth]{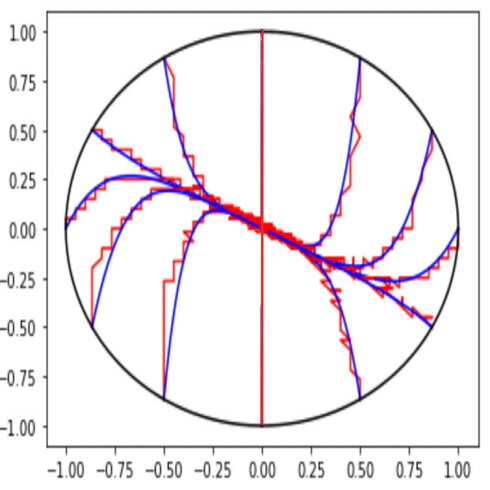}}
	\caption{Figure (a) shows the learned trajectories after 20 training episodes of a DQN model with randomly initialized parameters, while figure (b) uses the previously trained LTI system DQN model as initializations for the new system's model.}
	\label{fig:compare}
\end{figure}

To compare the training time of the model with initial weights and the existing model obtained from the previous emulation problem, we still choose the new emulated LTI system to be $\dot z=H_oz=\begin{bmatrix}
	-0.5& 0\\
	-1&-2.5
\end{bmatrix}z$. The existing DQN model is obtained from emulating the LTI system $\dot x=Hx=\begin{bmatrix}
	0& 1\\
	-1&-2
\end{bmatrix}x$, which generates the trajectory of Fig.\ref{fig:model}(a). Then, we train 20 episodes of these two models, respectively. Simulation results are shown in Fig. \ref{fig:compare}. It can be observed that the existing model has a better emulation performance after the same training episodes. The reason is that borrowing the existing model's structure and parameters for the new problem can reduce the value of the loss function to a certain extent during the initial training to shorten the training time.
\section{Conclusions and Future Work}
The work we have reported examines techniques in machine learning by which the neuromimetic linear models introduced in \cite{baillieul2021neuromimetic} and \cite{Sun2022} can learn to emulate stable closed-loop systems. We have proposed a learning model incorporating model predictive control (MPC) as well as a custom deep Q network (DQN) algorithm. For both approaches, we have studied specific cases in which transfer learning is possible for systems related by certain classes of coordinate transformations. For all cases, it is noted that the kinds of resilience to channel dropouts reported previously persist in the learned models. Future work will be aimed at showing how a continuous observer can be used to synthesize quantized trajectories for the class of overcomplete systems under consideration.

\bibliographystyle{IEEEtrans}

\begin{thebibliography}{3} 
	\bibitem{baillieul2014}
	J. Baillieul and Z. Kong, “Saliency based control in random feature networks,” in 53rd IEEE Conference on Decision and Control’, Los Angeles, CA, 2014, pp. 4210-4215. doi: 10.1109/CDC.2014.7040045
	
	\bibitem{Sutton2018}
	R. S. Sutton and A. G. Barto, Reinforcement learning: An introduction. MIT press, 2018.
	\bibitem{Lazaric2012}
	A. Lazaric, “Transfer in reinforcement learning: a framework and a survey.” in {\em Reinforcement Learning: State-of-the-Art}, Springer, 2012.
	\bibitem{bellman1957}
	R. Bellman, “A markovian decision process,” Journal of
	mathematics and mechanics, 1957.

	\bibitem{Arulkumaran2017}
	K. Arulkumaran, M. P. Deisenroth, M. Brundage, and A. A. Bharath, “A brief survey of deep reinforcement learning,” arXiv preprint arXiv:1708.05866, 2017.
	\bibitem{baillieul2019}
	J. Baillieul, "Perceptual Control with Large Feature and Actuator Networks," 2019 IEEE 58th Conference on Decision and Control (CDC), Nice, France, 2019, pp. 3819-3826, doi: 10.1109/CDC40024.2019.9029615.
	\bibitem{Zhu2020}
	Zhu, Z., Lin, K. and Zhou, J., "Transfer learning in deep reinforcement learning: A survey", arXiv preprint arXiv:2009.07888, 2020.
	\bibitem{Pan2009}
	S. J. Pan and Q. Yang, “A survey on transfer learning,” IEEE Transactions on knowledge and data engineering, 2009.
	\bibitem{baillieul2021neuromimetic}
	J. Baillieul and Z. Sun, “Neuromimetic Control–A Linear Model Paradigm,” 2021 60th IEEE Conference on Decision and Control (CDC), 2021, pp. 2709-2716, doi: 10.1109/CDC45484.2021.9683392.
	\bibitem{Sun2022}
	Z. Sun, and J. Baillieul. "Neuromimetic Linear Systems—Resilience and Learning." 2022 IEEE 61st Conference on Decision and Control (CDC), pp. 7388-7394. IEEE, 2022.
	
	\bibitem{DQN}
	Mnih, V., Kavukcuoglu, K., Silver, D. et al. Human-level control through deep reinforcement learning. Nature 518, 529–533 (2015). https://doi.org/10.1038/nature14236
	\bibitem{Hessel2018}
	Matteo Hessel, Joseph Modayil, Hado Van Hasselt, Tom Schaul, Georg Ostrovski, Will Dabney, Dan Horgan, Bilal Piot, Mohammad Azar, and David Silver. Rainbow: Combining improvements in deep reinforcement learning. In Thirty-Second AAAI Conference on Artificial Intelligence, 2018.
	\bibitem{tan2018}
	C. Tan, F. Sun, T. Kong, W. Zhang, C. Yang, and C. Liu. "A survey on deep transfer learning." In Artificial Neural Networks and Machine Learning–ICANN 2018: 27th International Conference on Artificial Neural Networks, Rhodes, Greece, October 4-7, 2018, Proceedings, Part III 27, pp. 270-279. Springer International Publishing, 2018.
	\bibitem{Tsitsiklis1997}
	John N Tsitsiklis and Benjamin Van Roy. An analysis of temporal-difference learning with function approximation. IEEE transactions on automatic control, 42(5):674–690, 1997.
	\bibitem{Van2018}
	Hado Van Hasselt, Yotam Doron, Florian Strub, Matteo Hessel, Nicolas Sonnerat, and Joseph Modayil. Deep reinforcement learning and the deadly triad. arXiv preprint arXiv:1812.02648, 2018.
	\bibitem{vehicle}
	S. El-Tantawy, B. Abdulhai, and H. Abdelgawad, “Multiagent reinforcement learning for integrated network of adaptive traffic signal controllers (marlin-atsc): methodology and large-scale application on downtown	toronto,” IEEE Transactions on Intelligent Transportation Systems, 2013.
	\bibitem{Sun2022MPC}
	Z. Sun, and J. Baillieul. "Model Predictive Control for Neuromimetic Quantized Systems." arXiv preprint arXiv:2212.09887, 2022.
	\bibitem{Oja1982}
	E. Oja, “Simplified neuron model as a principal component analyzer,” 1982. Journal of Mathematical Biology, 15(3), pp.267-273.
	\bibitem{cDQN}
	Wang, Z.T. and Ueda, M., 2021. Convergent and efficient deep Q network algorithm. arXiv preprint arXiv:2106.15419.
	\bibitem{Levine2018}
	Sergey Levine, Peter Pastor, Alex Krizhevsky, Julian Ibarz, and Deirdre Quillen. Learning hand- eye coordination for robotic grasping with deep learning and large-scale data collection. The International Journal of Robotics Research, 37(4-5):421–436, 2018.
	\bibitem{john1989}
	Christopher John Cornish Hellaby Watkins. Learning from delayed rewards. 1989.
\end{thebibliography}

\end{document}